\documentclass[superscriptaddress,preprint,nofootinbib]{article}
\pdfoutput=1 
\usepackage{jheppub}
\usepackage{graphicx}
\usepackage{amssymb}
\usepackage{amsmath}
\usepackage{epstopdf}
\usepackage{subfigure}
\usepackage{color}
\usepackage{mathtools}
\usepackage{bbm}
\usepackage{mathrsfs}
\usepackage{tikz}
\usepackage{pgfplots}
\usepackage[utf8]{inputenc}
\usepackage{todonotes}
\usepackage{enumerate}
\usepackage{gitinfo}

\pgfplotsset{every axis/.append style={
                    axis x line=middle,    
                    axis y line=middle,
                    xlabel = {$Z$},
                    ylabel = {$T$},
                    xtickmax = 0, xtickmin = 0,
                    ytickmax = 0, ytickmin = 0    
                    }}

\graphicspath{{./Artwork/}}

\def\kx{k\cdot x}
\def\eps{\varepsilon}
\def\dA{\dot{A}}
\def\dB{\dot{B}}
\def\dC{\dot{C}}
\def\dD{\dot{D}}

\def\wb{\bar{w}}

\def\pb{\bar{\partial}}

\def\be{\begin{equation}}
\def\ee{\end{equation}}
\def\bea{\begin{eqnarray}}
\def\eea{\end{eqnarray}}

\begin{document}

\subheader{\hfill \begin{tabular}{r} \texttt{QMUL-PH-18-27} \end{tabular}}

\title{Type D Spacetimes and the Weyl Double Copy}

\author[a]{Andr\'es Luna,}
\author[b]{Ricardo Monteiro,}
\author[c]{Isobel Nicholson,}
\author[c]{and Donal O'Connell}
\affiliation[a]{Mani L. Bhaumik Institute for Theoretical Physics, Department of Physics and Astronomy,
University of California at Los Angeles, California 90095, USA}
\affiliation[b]{Centre for Research in String Theory, School of Physics and Astronomy, Queen Mary University of London, 327 Mile End Road, London E1 4NS, UK}
\affiliation[b]{
Higgs Centre for Theoretical Physics, School of Physics and Astronomy, The University of Edinburgh, Edinburgh EH9 3JZ, Scotland, UK%
}
\emailAdd{ luna@physics.ucla.edu, ricardo.monteiro@qmul.ac.uk, \\ i.nicholson@sms.ed.ac.uk, donal@ph.ed.ac.uk}

\abstract{
We study the double-copy relation between classical solutions in gauge theory and gravity, focusing on four-dimensional vacuum metrics of algebraic type D, a class that includes several important solutions. We present a double copy of curvatures that applies to all spacetimes of this type -- the Weyl double copy -- relating the curvature of the spacetime to an electromagnetic field strength. We show that the Weyl double copy is consistent with the previously known Kerr-Schild double copy, and in fact resolves certain ambiguities of the latter. The most interesting new example of the classical double copy presented here is that of the C-metric. This well-known solution, which represents a pair of uniformly accelerated black holes, is mapped to the Li\'enard-Wiechert potential for a pair of uniformly accelerated charges. We also present a new double-copy interpretation of the Eguchi-Hanson instanton.
}

\maketitle

\section{Introduction}

There are close analogies between the Einstein equations and the Maxwell equations, and it is certainly very helpful to be introduced to the former only after becoming acquainted with the latter. There are also, however, obvious differences, one of them being that the Einstein equations are non-linear, whereas the Maxwell equations are linear---gravitons self-interact, but photons do not. A closer analogy, one may suspect, could be provided by the Yang-Mills equations, which are also non-linear. Indeed, there is a precise relation between Einstein's gravity and Yang-Mills theory, known as the {\it double copy}.

The double copy was discovered in the context of perturbation theory, in particular perturbative scattering amplitudes. The first step was taken by Kawai, Lewellen and Tye (KLT)~\cite{Kawai:1985xq} who demonstrated that any tree closed string amplitude can be expressed as a linear sum of factors, each of which is a product of two tree open string amplitudes. Since gauge interactions are described by open strings while  gravitational interactions are described by closed strings, the KLT relations in particular imply some kind of relationship between gauge theory and gravity. This relationship is a double copy, since to construct a gravitational amplitude, one needs to take products of two gauge theory amplitudes. 

It is possible to take the field theory limit of the full string-theoretic KLT relations, which leads to a direct set of relations between the scattering amplitudes of Yang-Mills theory and the amplitudes of the massless sector of string theory: this includes the usual graviton, as well as a scalar dilaton and an antisymmetric tensor. The focus of this article will be on the double copy in classical field theory. Moreover, we will restrict to pure Einstein gravity, so that the dilaton and the antisymmetric tensor are not present; we will comment on these fields in section~\ref{sec:discussion}.

The KLT relations have the advantage that an underlying reason for the relation between gauge theory and gravity is clear: by joining two open strings, you get one closed string. But they have the disadvantage that the relations themselves become quite complicated for high multiplicity. More recently, Bern, Carrasco and Johansson (BCJ) discovered a new and simple form for the double copy~\cite{Bern:2008qj} which also leads to a natural formulation of the double copy at loop level~\cite{Bern:2010ue}. This form of the double copy has been extensively studied at tree level, leading to various proofs~\cite{BjerrumBohr:2009rd,Stieberger:2009hq,Bern:2010yg,BjerrumBohr:2010yc,Chen:2010ct,Chen:2011jxa,Mafra:2011kj,BjerrumBohr:2012mg,Cachazo:2012uq,Tolotti:2013caa,Monteiro:2013rya,Bjerrum-Bohr:2016axv,Du:2017kpo}. At loop level, the double copy has been extensively studied~\cite{Carrasco:2011mn,Boels:2012ew,Boels:2013bi,Bern:2013yya,Bjerrum-Bohr:2013iza,Johansson:2014zca,Mafra:2015mja,He:2015wgf,Mogull:2015adi,Yang:2016ear,Bern:2017yxu,Johansson:2017bfl,He:2016mzd,He:2017spx,Geyer:2017ela}, but to date it is still a conjecture~\cite{Bern:2017yxu}. Nevertheless, the BCJ double copy is a powerful tool in the theory of scattering amplitudes, which has led to rich new insights into the structure of supergravity, e.g.~\cite{Anastasiou:2014qba,Anastasiou:2016csv,Chiodaroli:2017ngp,Bern:2017tuc,Anastasiou:2017nsz,Bern:2017ucb,Chiodaroli:2017ehv,Bern:2017rjw,Anastasiou:2018rdx}. A celebrated recent example is the detailed computation of the UV structure of maximal supergravity at five loops~\cite{Bern:2018jmv}. The double copy is reviewed in, for example,~\cite{Carrasco:2015iwa,Chiodaroli:2016jqw,Cheung:2017pzi}.

The success of the double copy for scattering amplitudes motivated the investigation of its manifestation for solutions of the classical field equations, with early steps given in~\cite{Bern:1999ji, Saotome:2012vy, Neill:2013wsa}. Since the principles of the double copy, as currently understood, are perturbative in nature, it is remarkable that exact relations between solutions can be found, as discovered in \cite{Monteiro:2014cda}. In the same way that solutions to the Maxwell equations provide a class of linear solutions to the Yang-Mills equations (with trivial colour dependence), there is a class of solutions that linearise the Einstein equations. Kerr-Schild metrics belong to this class, and so do certain multi-Kerr-Schild metrics. It is natural, therefore, to study the double copy in this context, and a {\it Kerr-Schild double copy} was found in \cite{Monteiro:2014cda}. A particularly interesting example that it applies to is the Kerr-Taub-NUT family of solutions \cite{Luna:2015paa}. The self-dual sector of gauge theory and gravity is another natural setting to study the double copy, as found in~\cite{Monteiro:2011pc} at the perturbative level. Very recently, the Eguchi-Hanson metric was studied in this context using the Kerr-Schild double copy~\cite{Berman:2018hwd}. For other work on the Kerr-Schild double copy, see~\cite{Ridgway:2015fdl,Luna:2016due,White:2016jzc,Adamo:2017nia,DeSmet:2017rve,Bahjat-Abbas:2017htu,Carrillo-Gonzalez:2017iyj,Ilderton:2018lsf,Lee:2018gxc,Gurses:2018ckx}.

In this paper, we will introduce a different type of classical double copy, one that involves curvatures (the Weyl curvature and the Maxwell field strength) rather than fields, and which we call the {\it Weyl double copy}. It applies to any four-dimensional vacuum spacetime of algebraic type D, which includes not only the Kerr-Taub-NUT family, for which the Weyl double copy reproduces the results from the Kerr-Schild double copy, but also solutions with an acceleration parameter such as the C-metric. The relation to Maxwell solutions follows from the (complex) double-Kerr-Schild structure of type D solutions \cite{Plebanski:1976gy}. As we shall discuss, there is a distinction between the existence of (multi-)Kerr-Schild coordinates and the applicability of the Kerr-Schild double copy as prescribed in \cite{Monteiro:2014cda}; the former is necessary but not sufficient for the latter. It turns out that the Weyl double copy was partly anticipated in \cite{Walker:1970un,Hughston:1972qf}, where the existence of a Killing spinor for type D spacetimes was established. This Killing spinor underlies the Weyl double copy, as we will explain.

We consider other solutions of interest. For pp-waves, the Weyl double copy resolves an ambiguity in the Kerr-Schild double-copy procedure, by picking up a unique and very natural correspondence between gravitational and gauge theory wave solutions. We study also the Eguchi-Hanson instanton, as an interesting example of a self-dual spacetime. We find that the straightforward Weyl double-copy interpretation of the solution differs from that given in \cite{Berman:2018hwd}. The results of \cite{Berman:2018hwd} are reinterpreted as a `mixed' double copy involving a pair of distinct gauge theory solutions.

While we focus on four-dimensional spacetimes, we expect that the double copy of curvatures exploited here extends to higher dimensions. In view of this goal, ref.~\cite{Monteiro:2018xev} has recently revisited the problem of extending to higher dimensions the well-known spinorial formalism that we employ here. 

We emphasise that the double copy relates Einstein's gravity and Yang-Mills theory, and that generic gravitational solutions cannot be related to Maxwell theory. If gravitational solutions can be written in closed form, however, that is because they possess a large amount of symmetry (hidden or not), and it is not surprising that there may be an underlying linear structure. For generic solutions, there is ample evidence that the double copy applies in perturbation theory~\cite{Cardoso:2016ngt,Goldberger:2016iau,Cardoso:2016amd,Luna:2016hge,Cheung:2016say,Cheung:2017kzx,Goldberger:2017frp,Luna:2017dtq,Goldberger:2017vcg,Chester:2017vcz,Goldberger:2017ogt,Li:2018qap,Shen:2018ebu,Plefka:2018dpa,Mizera:2018jbh,Carrillo-Gonzalez:2018pjk}, but its exact (non-perturbative) formulation remains elusive. 

This paper is organised as follows. Section~\ref{sec:reviewKS} is a review of the Kerr-Schild double copy. In section~\ref{sec:idea}, we introduce the Weyl double copy, and give basic examples. The application of the Weyl double copy to the vacuum type D family of solutions is discussed in section~\ref{sec:generaltypeD}, at the end of which the example of the C-metric is presented in detail. The Eguchi-Hanson instanton is studied in section~\ref{sec:EH}. Finally, we discuss possible future directions in section~\ref{sec:discussion}.

\section{Review of the Kerr-Schild double copy}
\label{sec:reviewKS}

The Kerr-Schild double copy relates a class of Kerr-Schild spacetimes, to be discussed below, to solutions of the Maxwell equations \cite{Monteiro:2014cda}. There are two properties of Kerr-Schild spacetimes that make them ideally suited to the classical double copy:
\begin{enumerate}
\item Kerr-Schild spacetimes can be expressed as a deviation from a base spacetime, which we take here to be Minkowski spacetime. Therefore, they possess a natural set of (Kerr-Schild) coordinates that map trivially to the flat spacetime in which the gauge theory lives, and this makes the double copy between gauge field and metric much simpler. 
\item The Kerr-Schild ansatz linearises the Einstein equations. Therefore, it makes sense to associate this type of spacetime to Abelian gauge field configurations, i.e., to solutions of the Maxwell equations. 
\end{enumerate}
The defining property of Kerr-Schild spacetimes is that they admit coordinates for which the metric components read
\begin{equation}
\label{KSmetric}
g_{\mu\nu}=\eta_{\mu\nu}+ \phi\, k_\mu k_\nu \,,
\end{equation}
where $k_\mu$ is null and geodesic with respect to the Minkowski metric $\eta_{\mu\nu}$. It is easy to show that $k_\mu$ is then also null and geodesic with respect to the curved metric $g_{\mu\nu}$. The spacetime can be thought of as a deviation from Minkowski spacetime.\footnote{There are caveats to this interpretation \cite{Gibbons:2017djb}, but they are not crucial for the present discussion.} Notice that the inverse metric takes the simple form
\begin{equation}
g^{\mu\nu}=\eta^{\mu\nu}-{ \phi\, k^\mu k^\nu} \,, \qquad \text{with} \quad k^\mu=g^{\mu\nu}k_\nu=\eta^{\mu\nu}k_\nu\,.
\end{equation}
What makes the Kerr-Schild form of the metric famous is that it linearises the Ricci tensor with mixed indices:
\begin{equation}
\label{RudKS}
R^\mu{}_\nu=\frac{1}{2}  \partial_\alpha \left[\partial^\mu \left({ \phi k^\alpha k_\nu}\right)+\partial_\nu \left({ \phi k^\alpha k^\mu}\right)-\partial^\alpha \left({ \phi k^\mu k_\nu}\right)\right], \qquad \partial^\mu\equiv \eta^{\mu\nu} \partial_\nu \,.
\end{equation}

Let us take $\,\eta_{\mu\nu}=\textrm{diag}(-1,1,\cdots,1)\,$, and suppose that the metric is time independent, so that $\partial_0\phi=0$ and $\partial_0k_\mu=0$. We will refer to such a spacetime as a stationary Kerr-Schild spacetime. We also choose to set $k_0=1$.\footnote{This is achieved via the rescaling $(\phi,k_\mu) \to (\phi \,k_0^{\,2},k_\mu/k_0)$. Of course, the rescaled $k_\mu$ is not necessarily geodesic, but it is still such that \eqref{RudKS} follows from \eqref{KSmetric}.} The vacuum Einstein equations read 
\begin{equation}
\label{KSeqs}
R^{0}{}_{0}=\frac{1}{2}{\partial^i\partial_i \phi} =0\,, \qquad \qquad
R^{i}{}_0=\frac{1}{2}{ \partial_j \left[\partial^i\left(\phi k^j\right)-\partial^j\left(\phi k^i\right)\right]}=0\,,
\end{equation}
together with the remaining $R^{i}{}_j=0$. Now, the equations \eqref{KSeqs} coincide precisely with the Maxwell equations
\begin{equation}
\partial^\mu F_{\mu0}={\partial^i\partial_i \phi} =0\,, \qquad \qquad
\partial^\mu F_{\mu i}={ \partial^j \left[\partial_j\left(\phi k_i\right)-\partial_i\left(\phi k_j\right)\right]}=0\,,
\end{equation}
for the gauge field 
\begin{equation}
\label{KSA}
A_\mu = \phi \, k_\mu \,.
\end{equation}

The relation between the gauge field \eqref{KSA} and the metric \eqref{KSmetric} is the {\it Kerr-Schild double copy}. The double-copy interpretation is supported by the analogy with the double copy for scattering amplitudes and by the perturbative construction of the double copy for classical solutions; see \cite{Monteiro:2014cda} and \cite{Luna:2016hge} for discussions.

Notice that we didn't discuss how the spatial components of the Einstein equations, $R^{i}{}_j=0$, relate to the gauge theory solution. They represent a constraint on the latter. This is analogous to the situation in scattering amplitudes, where the kinematic numerators of the gauge theory amplitude must satisfy colour-kinematics duality for the double copy to hold \cite{Bern:2008qj}. It is interesting to note that this constraint is a three-term identity, just like the Jacobi relation in the colour-kinematics duality. 

Let us look at the simplest example of the Kerr-Schild double copy. Consider the four-dimensional solutions with
\begin{equation}
\phi(r) = \frac{C}{r},\qquad k_\mu=\left(1,\frac{\vec{x}}{r}\,\right)\,,
\end{equation}
where $r^2=x_ix^i$. In the gravity case, we have the Schwarzschild solution with $C=2M$, where $M$ is the mass. In the gauge theory case, we have the Coulomb solution with $C$ representing the charge; this becomes more obvious after a gauge transformation,
\begin{equation}
A_\mu=\phi\,k_\mu \quad \longrightarrow \quad A'_\mu=\frac{C}{r} \,(1,\vec{0})\,.
\end{equation}
In any sensible definition of the classical double copy, these two spherically-symmetric, static solutions should be related, and this is indeed achieved. There are many other examples in the literature, some of which will be reproduced using the Weyl double copy defined in this paper.

Before proceeding, we point out that the beautiful properties of Kerr-Schild spacetimes extend to a class of multi-Kerr-Schild spacetimes. For instance, double-Kerr-Schild spacetimes admit a metric of the type
\begin{equation}
\label{doubleKSmetric}
g_{\mu\nu}=\eta_{\mu\nu}+ \phi\, k_\mu k_\nu + \psi\, \ell_\mu \ell_\nu \,,
\end{equation}
where $k_\mu$ and $\ell_\mu$ are null, geodesic and mutually orthogonal. The linearisation property \eqref{RudKS} generically fails for multi-Kerr-Schild spacetimes, but there are exceptions of interest. For example, certain double-Kerr-Schild spacetimes are such that
\begin{equation}
\label{doubleRudKS}
R^\mu{}_\nu[ \phi\, k_\mu k_\nu + \psi\, \ell_\mu \ell_\nu]=R^\mu{}_\nu[\phi\, k_\mu k_\nu] + R^\mu{}_\nu[\psi\, \ell_\mu \ell_\nu] \,.
\end{equation}
We will encounter examples in section~\ref{sec:generaltypeD}.

\section{The Weyl double copy}
\label{sec:idea}

As a motivation, let us consider linear waves in gauge theory and in gravity:
\begin{align}
& A_\mu = \epsilon_\mu\, e^{i\kx}\,, \quad F_{\mu\nu}=i(k_\mu\epsilon_\nu-k_\nu\epsilon_\mu)\, e^{i\kx}\,, \nonumber\\
& h_{\mu\nu} = \varepsilon_{\mu\nu}\, e^{i\kx}\,, \quad R_{\mu\nu\rho\lambda}=\frac{1}{2}(k_\mu\epsilon_\nu-k_\nu\epsilon_\mu)(k_\rho\epsilon_\lambda-k_\lambda\epsilon_\rho)\, e^{i\kx}\,, \quad \text{if} \;  \varepsilon_{\mu\nu}=\epsilon_\mu\epsilon_\nu \,.
\end{align}
There is an obvious double-copy relationship between the basic gauge-invariant quantities at linearised level,
\be
e^{i\kx} R_{\mu\nu\rho\lambda} \sim F_{\mu\nu} F_{\rho\lambda}.
\ee
Our goal is to explore this type of relationship for exact solutions, rather than linearised ones. The challenge is to match both the symmetries and the gauges in both sides of the relation. Kerr-Schild coordinates will play an important role in matching the gauges. As for the symmetries, the algebraic structure of the Riemann curvature is much simpler in a spinorial formalism, particularly in four dimensions, and this is the path that we will follow. Along the way, we will see that the double copy gives a fresh insight into basic results in general relativity.

\subsection{Spinorial formalism}
\label{sec:spinorial}

Our starting point is the spinorial formalism of general relativity \cite{Penrose:1960eq}. We start with the object $\sigma^\mu_{A\dA}$, the `spinorial vierbein', such that
\be
\left(\sigma^\mu_{A\dA} \sigma^\nu_{B\dB}+\sigma^\nu_{A\dA} \sigma^\mu_{B\dB}\right) \eps^{\dA\dB} = g^{\mu\nu} \eps_{AB}.
\ee
In our convention, $\eps^{12}=1$\,.
We can easily write
\be
\sigma^\mu_{A\dA} = (e^{-1})^\mu_a \,\sigma^a_{A\dA}, \qquad \sigma^a=\frac{1}{\sqrt{2}}(\mathbbm{1},\sigma^i),
\ee
where $a$ is a tangent space index, $\sigma^i$ are the Pauli matrices, and $(e^{-1})^\mu_a$ is the (inverse) vierbein, defined such that
\be
g^{\mu\nu} = (e^{-1})^\mu_a\, (e^{-1})^\nu_b\, \eta^{ab}, \qquad \eta^{ab}=\eta_{ab}=\text{diag}(-1,1,1,1).
\ee
Using $\sigma^\mu_{A\dA}$ and its inverse, which satisfy
\be
\sigma_\mu^{A\dA} = g_{\mu\nu} \,\eps^{AB}\,\sigma^\nu_{B\dB}\,\eps^{\dB\dA},
\qquad \sigma^\mu_{A\dA} \sigma_\nu^{A\dA} =\delta^\mu_\nu,
\qquad \sigma^\mu_{A\dA} \sigma_\mu^{B\dB} = \delta^B_A \delta^{\dB}_{\dA},
\ee
we can write any tensor in spinorial form. For instance, $V_\mu \mapsto V_{A\dA} = \sigma^\mu_{A\dA}\,V_\mu $.

We are interested in looking at the curvature. Let us focus on vacuum spacetimes, $R_{\mu\nu}=0$. The Riemann tensor then coincides with the Weyl tensor, $R_{\mu\nu\rho\lambda}=W_{\mu\nu\rho\lambda}$. The spinorial form of the Weyl tensor is
\be
\label{eq:Wspinor}
W_{A\dA B\dB C\dC D\dD} = C_{ABCD} \,\eps_{\dA\dB} \eps_{\dC\dD} + \bar{C}_{\dA\dB\dC\dD} \,\eps_{AB} \eps_{CD},
\ee
where $C_{ABCD}$ and $\bar{C}_{\dA\dB\dC\dD}$ are completely symmetric, and are related by complex conjugation if the (Lorentzian) spacetime is real. In fact, $C_{ABCD}$ and $\bar{C}_{\dA\dB\dC\dD}$ represent the anti-self-dual and self-dual parts of the curvature. We can obtain $C_{ABCD}$ from the curvature tensor directly as
\be
C_{ABCD} = \frac{1}{4} W_{\mu\nu\rho\lambda}\,\sigma^{\mu\nu}_{AB}\,\sigma^{\rho\lambda}_{CD},
\ee
using the object
\be
\sigma^{\mu\nu}_{AB} = \sigma^{[\mu}_{A\dA} \,{\tilde{\sigma}}^{\nu]\; \dA C}\, \eps_{CB},
\qquad {\tilde{\sigma}}^{\mu\; \dA A} =(e^{-1})^\mu_a\, {\tilde\sigma}^{a\;\dA A}\,,
\qquad {\tilde\sigma}^a=\frac{1}{\sqrt{2}}(\mathbbm{1},-\sigma^i).
\ee

For a gauge theory field strength $F_{\mu\nu}$, we have analogously
\be
\label{eq:Fspinor}
F_{A\dA B\dB} = f_{AB} \,\eps_{\dA\dB} + \bar{f}_{\dA\dB} \,\eps_{AB}\,, \qquad \text{with} \qquad
f_{AB} = \frac{1}{2} F_{\mu\nu}\, \sigma^{\mu\nu}_{AB}
\ee
and $f_{AB}=f_{BA}$, as well as $f_{AB}=(\bar{f}_{\dA\dB})^*$ if the field strength is real. The spinorial structure has a clear analogy with the gravitational case, but we will see that it is more than an analogy.

The key idea of this paper is a relation between exact solutions in gravity and in (flat spacetime) gauge theory, the {\it Weyl double copy}, which we define as
\be
\label{eq:Wdoublecopy}
\boxed{C_{ABCD} = \frac{1}{S}\, f_{(AB}\,f_{CD)}}\,.
\ee
As we will see, the scalar $S$ and the field strength spinor $f_{AB}$ are uniquely determined by the Weyl spinor of the gravity solution.
In the examples of single-Kerr-Schild spacetimes \eqref{KSmetric} that we will consider, the real part of $S$ coincides with $\phi$ up to a constant factor.

The  Weyl double copy \eqref{eq:Wdoublecopy} is related to the algebraic classification of spacetimes.\footnote{See \cite{Stephani:2003tm} or \cite{Griffiths:2009dfa} for more details on this classification.} In particular, it implies that the spacetime has Petrov type D or N, as we shall now see. Let us first notice that we can always decompose the Weyl spinor into four rank-1 spinors,
\be
C_{ABCD} = \mathbf{a}_{(A} \,\mathbf{b}_B \,\mathbf{c}_C \,\mathbf{d}_{D)}.
\ee
These four spinors give us the four principal null directions of the spacetime, e.g.,~$a_{A\dA}={\mathbf{a}}_A \,\bar{\mathbf{a}}_{\dA}$. The algebraic classification is based on whether the principal null directions coincide (up to scaling). If spacetimes have four distinct principal null directions, they are of type I (algebraically general), otherwise they are algebraically special. If all principal null directions coincide, i.e., there is a single principal null direction with multiplicity four, then the spacetime is of type N. If there are two principal null directions with multiplicity two, then it is of type D.\footnote{The possible types are I, II, III, D, N and O. The types that we did not mention before are: type II (three principal null directions, one of which with multiplicity two), type III (two principal null directions, one of which with multiplicity three), and type O (vanishing Weyl tensor).} Now, we can do the same for the field strength spinor,
\be
f_{AB} = \mathbf{r}_{(A}\, \mathbf{s}_{B)},
\ee
and also discuss principal null directions in this context -- there exist only two. Therefore, the field strength is algebraically general if it has two distinct principal null directions, and algebraically special if it has a single principal null direction with multiplicity two. It is then clear that a Weyl tensor satisfying \eqref{eq:Wdoublecopy} corresponds to a type D spacetime, if the field strength is general, or to a type N spacetime, if the field strength is special. In either case, the spacetime is algebraically special. We will see examples of both type N and type D. While the Weyl double copy studied in this paper applies only to type D and (certain) type N spacetimes, we emphasise that we expect a more general double-copy relation to exist.

Our proposal of the Weyl double copy was partly anticipated in the general relativity literature.\footnote{We thank Lionel Mason for bringing these references to our attention.} In \cite{Walker:1970un}, Penrose and Walker found that any type D spacetime admits a Killing rank-2 spinor $\chi_{AB}$, i.e., $\nabla_{(A}^{\;\;\;\;\dA}\;\chi_{BC)}=0$, and that the spacetime's Weyl spinor can be written as
\be
\label{eq:penrosewalker}
C_{ABCD} = [\chi]^{-5} \chi_{(AB}\,\chi_{CD)}\,, \qquad \text{with}\quad [{\chi}]=(\chi_{AB}\chi^{AB})^{1/2}\,.
\ee
Moreover, the same authors together with Hughston and Sommers pointed out in \cite{Hughston:1972qf} that the field strength spinor
\be\label{eq:ChiFieldStrength}
{\breve f}_{AB} = [\chi]^{-3} \chi_{AB} 
\ee
is a solution of the Maxwell equations on the curved background with Weyl spinor $C_{ABCD}$.
Therefore, a relation like \eqref{eq:Wdoublecopy} holds once we identify the scalar
\begin{equation}
\breve S = [\chi]^{-1}.
\end{equation}
This scalar satisfies an equation $\square \breve S \propto [\chi]^{-3}$ on the same background spacetime with curvature $C_{ABCD}$.
As we will see, these statements for type D spacetimes have natural analogues for pp-waves, a type N class of spacetimes representing exact waves \cite{Dietz361}.

To connect this more directly to the standard double copy, and to fully specify our Weyl double copy eq.~\eqref{eq:Wdoublecopy}, we point out that it is in fact possible in general to demand that \eqref{eq:Wdoublecopy} is valid with the fields on the right-hand-side, $ f_{AB}$ and $ S$, solving the Maxwell and wave equations of motion on a flat background, respectively.\footnote{Of course, it may be instructive to consider the double copy on a curved background as in \cite{Bahjat-Abbas:2017htu,Carrillo-Gonzalez:2017iyj}. This is not the intent of our work.} These flat-spacetime fields are easily obtained from the curved-spacetime fields using (double-)Kerr-Schild coordinates, as we shall show explicitly for general type D spacetimes. The relation \eqref{eq:Wdoublecopy} is then guaranteed to hold. In other words, given any type D spacetime (and any pp-wave), we can construct a gauge field and a scalar satisfying flat-spacetime equations of motion uniquely up to constant factors, and -- crucially -- satisfying the Weyl double copy. 

There are extensions of the Killing spinor structure to higher-dimensional spacetimes; see e.g. \cite{Frolov:2017kze,Mason:2010zzc}. 


\subsection{Basic examples}

We start with two basic examples. The first example is that of pp-waves, which are type N spacetimes representing exact wave solutions. This is an interesting example because the original double copy for scattering amplitudes relies on the double copy of perturbative wave solutions. We will find that the Weyl double copy resolves an ambiguity with the Kerr-Schild double copy. As a type D example, we will discuss how the Weyl double copy reproduces the Kerr-Schild double copy in the case of the Kerr spacetime.

\subsubsection{Exact wave solutions}
\label{sec:planewave}

The first example that we will consider is that of exact waves. In gravity, we have the pp-wave metric,
\be
ds^2= -dt^2+dx^2+dy^2+dz^2+h(t-z,x,y) (dt-dz)^2 = -du \,dv +dw \,d\wb +h(u,w,\wb) du^2,
\ee
where $u=t-z$, $v=t+z$, and $w=x+iy$.
For the vierbein, we can take
\be
e^a_{\phantom{a}\mu} =\frac{1}{2} \left( \begin{array}{cccc}
1-h & 1 & 0 & 0 \\
1+h & -1 & 0 & 0 \\
0 & 0 & 1 & 1 \\
0 & 0 & i & -i \end{array} \right).
\ee
The equation of motion is simply $\partial \pb h(u,w,\wb)=0$, where we use $\partial\equiv\partial_w$, $\pb\equiv\partial_{\wb}$. Using an arbitrary spinor $\xi^A=(\xi^1,\xi^2)$, we can conveniently express the Weyl spinor as the polynomial
\be
C_{ABCD} \,\xi^A\xi^B\xi^C\xi^D = \frac{1}{2}(\xi^1+\xi^2)^4\, \partial^2 h.
\ee
This spacetime is of type N, since there is a single principal null direction with multiplicity four. The combination $\xi^1+\xi^2$ is associated with $a_{A\dA}={\mathbf{a}}_A \,\bar{\mathbf{a}}_{\dA}$ in the case ${\mathbf{a}}_A=(1,1)=\bar{\mathbf{a}}_{\dA}$, so that $a_\mu = \sigma_\mu^{A\dA} \,a_{A\dA} =\sqrt{2}\, \delta_\mu^u $. Hence, the principal null direction is given by the covector $du$ or, as a vector, by $\partial_v$, which is the Killing vector field (i.e., the line element $ds^2$ does not depend on translations of $v$).

Even without considering any double copy, it is easy to construct an analogous gauge theory solution, which is
\be
A = {\mathcal A}(u,w,\wb) \,du, \qquad \partial \pb {\mathcal A}(u,w,\wb)=0.
\ee
The associated field strength spinor is simply
\be
f_{AB} \,\xi^A\xi^B = -i(\xi^1+\xi^2)^2\, \partial {\mathcal A}.
\ee

Now let us consider these two solutions from the point of view of the Weyl double copy. Evidently equation~\eqref{eq:Wdoublecopy} holds provided we identify
\be
\label{eq:Swave}
S = -2 \frac{(\partial {\mathcal A})^2}{\partial^2 h}. 
\ee
It is clear that $S$ satisfies the wave equation, since \,$\partial \pb S(u,w,\wb)=0$\,. In fact, it satisfies the stronger condition $\pb S(u,w,\wb)=0$\,.

This example raises an important question for the classical double copy: to what extent is it unique? In the stationary Kerr-Schild case, there is a natural, unique choice of scalar. But in these wave examples,
it appears that any element of a class of vacuum gravity solutions (i.e., any $h$ satisfying $\partial \pb h=0$) can be related to any element of a class of gauge theory solutions (i.e., any ${\mathcal A}$ satisfying $\partial \pb {\mathcal A}=0$), given an appropriate choice of the scalar field $S$, namely eq. \eqref{eq:Swave}.\footnote{We are restricting ourselves presently to vacuum gravity solutions. For the reader familiar with the double copy literature, we point out that the issue of non-uniqueness here is distinct from the question of a more general double copy involving the dilaton and the two-form field on the gravity side. We briefly refer to this question in the Discussion section.} So from this point of view it seems that the double copy is non-unique.

Indeed, the authors of Ref.~\cite{Monteiro:2014cda} discussed exact waves briefly while introducing the Kerr-Schild double copy. These authors
pointed out that the functions $h$ and ${\mathcal A}$ appearing in the gravity and the gauge theory cases, respectively, satisfy the same equation of motion. Now, it is obviously tempting to identify the functions $h$ and ${\mathcal A}$ via the double copy. But if this identification is made, then a \textit{plane} wave in gravity does not map to a {\it plane} wave in gauge theory. It maps instead to a vortex solution~\cite{Ilderton:2018lsf}. 

Nevertheless, the expectation of a double copy of plane waves to plane waves is very natural and is supported by the results of Ref.~\cite{Adamo:2017nia}, where it was shown that an appropriately defined double copy of gluon scattering amplitudes on a gauge theory plane wave background is a gravity scattering amplitude on a gravitational plane wave background, at least for three-point scattering; see \cite{Adamo:2018mpq} for recent advances beyond three points. 

Although it may be of interest to explore more general notions of the double copy, the Weyl double copy we are proposing settles this issue in favour of uniqueness.  Let us recall the discussion of equations \eqref{eq:penrosewalker} and \eqref{eq:ChiFieldStrength}, which applies to pp-waves as well as to type D spacetimes.
We can see that $C_{ABCD}\sim [\chi]^{-3}\sim \partial^2h$, so that $[\chi]=(\partial^2 h)^{-1/3}$. For the gauge field, $f_{AB}\sim [\chi]^{-2}\sim \partial {\mathcal A}$. Finally, the scalar field $S$ is identified with $[\chi]^{-1}$ and, therefore, with $(\partial^2 h)^{1/3}$.

In particular, in the plane wave case the functions $h$ and $\mathcal A$ are
\begin{align}
h &= {\mathfrak h}(u)\,w^2 + \bar{{\mathfrak h}}(u)\,\wb^2 \,, \\
{\mathcal A} &= {\mathfrak a}(u)\,w + \bar{{\mathfrak a}}(u)\,\wb \,.
\end{align}
In this case, the Weyl double copy requires $[\chi] = {\mathfrak h}^{-1/3}$ and ${\mathfrak a}(u)={\mathfrak h}^{2/3}$.

We will see in section~\ref{sec:generaltypeD} that, for general type D vacuum solutions, this unique procedure to identify the single copy leads to a very natural result.

\subsubsection{Kerr solution}
\label{sec:kerr}

The Kerr metric, possibly the most important exact solution in general relativity, represents an asymptotically flat rotating black hole for $|a|\leq M$, where $a$ is the rotation parameter and $M$ is the mass. In Boyer-Lindquist coordinates, but using $X=\cos\theta$, the metric is
\be
\label{eq:kerrmetric}
ds^2 = -dt^2 + \Sigma^2 \left( \frac{dr^2}{\Delta} +\frac{dX^2}{1-X^2} \right)+(r^2+a^2)(1-X^2) d\psi^2
+ \frac{2Mr}{\Sigma^2}\left(dt+a(1-X^2)d\psi\right)^2,
\ee
where
\be
\Sigma^2=r^2+a^2X^2, \qquad \Delta=r^2+a^2-2Mr.
\ee
The vierbein can be given by
\be
e^a_{\phantom{a}\mu} = \left( \begin{array}{cccc}
\sqrt{\Delta/\Sigma^2} & 0 & 0 & a(1-X^2)\sqrt{\Delta/\Sigma^2} \\
0 & \sqrt{\Sigma^2/\Delta} & 0 & 0 \\
0 & 0 & \sqrt{\Sigma^2/(1-X^2)} & 0 \\
a \sqrt{(1-X^2)/\Sigma^2} & 0 & 0 & (r^2+a^2) \sqrt{(1-X^2)/\Sigma^2}\end{array} \right).
\ee
The Weyl spinor is
\be
C_{ABCD} \,\xi^A\xi^B\xi^C\xi^D = -\frac{3M}{2(r+iaX)^3}\left((\xi^1)^2-(\xi^2)^2\right)^2.
\ee
It clearly shows that this is a type D spacetime, since there are two principal null directions, each with multiplicity two. These correspond to the two combinations $\xi^1\pm\xi^2$.

According to equation \eqref{eq:ChiFieldStrength}, we would expect a Maxwell field of the form
\be
\breve f_{AB}\, \xi^A\xi^B \propto \frac{1}{(r+iaX)^2}\left((\xi^1)^2-(\xi^2)^2\right).
\ee
to live on the curved spacetime \eqref{eq:kerrmetric}. This is indeed the case. However, the standard double copy is expected to relate the dynamics of gauge theory to the dynamics of gravity, so we would like to construct a gauge field $f_{AB}$ living on flat spacetime. This is straightforward: we note that 
\be\label{eq:kerrFieldStrength}
f_{AB}\, \xi^A\xi^B = -\frac{Q}{2(r+iaX)^2}\left((\xi^1)^2-(\xi^2)^2\right)
\ee
is a Maxwell solution which lives on flat spacetime, with a metric given as in \eqref{eq:kerrmetric} but with $M=0$\footnote{We use in the following a ``flat'' superscript $k^{(0)}_\mu$ because the gauge theory solution is a solution in Minkowski spacetime, and therefore one has to take $M=0$ in the Kerr-Schild covector written in Boyer-Lindquist coordinates; this is not required in the Kerr-Schild coordinates used in \cite{Monteiro:2014cda}, for which the curved spacetime / flat spacetime map is trivial. For an overview of common coordinate systems for the Kerr solution, see e.g.~\cite{Visser:2007fj}.}. The Kerr solution has already been analysed in the context of the double copy \cite{Monteiro:2014cda} and the gauge field was found to be 
\be
A_\mu = \frac{Q\, r}{\Sigma^2} \,k^{(0)}_\mu, \qquad k^{(0)}=dt+\frac{\Sigma^2}{r^2+a^2} dr+a(1-X^2) d\psi,
\ee
which leads to the correct field strength \eqref{eq:kerrFieldStrength}.

According to the Weyl double copy \eqref{eq:Wdoublecopy}, we have
\be
S = -\frac{Q^2}{6M} \,\frac{1}{r+iaX}.
\ee
This complex function satisfies the wave equation in Minkowski spacetime. The real scalar is given by
\be
S+\bar{S}= -\frac{Q^2}{3M} \,\frac{r}{\Sigma^2}.
\ee
This is precisely (up to a different normalisation constant) the Kerr-Schild scalar $\phi$ identified in \cite{Monteiro:2014cda}, which plays the role of the ``zeroth copy''. Note that equations \eqref{eq:Wspinor} and \eqref{eq:Fspinor} illuminate the different roles of $S$ and $\phi$: $\phi$ is comparable to the Weyl and field strength tensors,  while $S$ is comparable to the complex spinor objects $C_{ABCD}$ and $f_{AB}$.


\section{General vacuum type D metrics}
\label{sec:generaltypeD}

The Plebanski-Demianski family\footnote{This family of solutions was first found by Debever \cite{Debever:1971}, but is best known in the Plebanski-Demianski form. While our interest lies in the case with $\Lambda = 0$ and vanishing Maxwell field, the Plebanski-Demianski family also includes a large class of type D solutions of the Einstein-Maxwell equations with a cosmological constant.} of metrics~\cite{Plebanski:1976gy} describes the most general vacuum type D solution of the Einstein equations (for vanishing Maxwell field.)
We will now study this family of solutions from the perspective of the Weyl double copy, restricting ourselves to the vanishing cosmological constant case, which has a simpler double-copy interpretation. We construct the electromagnetic and scalar fields associated with this family. In section~\ref{sec:scaling}, we will describe two particularly interesting examples of this family, beyond the Kerr example seen previously: the Kerr-Taub-NUT solution, whose double-copy structure was studied in \cite{Luna:2015paa}, and the C-metric, which receives here its first double-copy analysis. A useful reference when considering various coordinate systems of the Plebanski-Demianski family is \cite{Griffiths:2005qp}.

\subsection{Double copy}

The general vacuum type D solution with vanishing cosmological constant takes the form \cite{Plebanski:1976gy}
\begin{align}
ds^2 = \frac{1}{(1-pq)^2} \left[ \frac{p^2+q^2}{P(p)}\,dp^2 +  \frac{P(p)}{p^2+q^2}\,(d\tau+q^2d\sigma)^2 + \frac{p^2+q^2}{Q(q)}\,dq^2 -  \frac{Q(q)}{p^2+q^2}\,(d\tau-p^2d\sigma)^2 \right]\,,
\end{align}
with 
\begin{align}
P(p) =  \gamma (1-p^4) +2 n p- \epsilon p^2 +2 m p^3 \,, \nonumber \\
Q(q) =  \gamma (1-q^4) - 2m q + \epsilon q^2 -2 n q^3 \,,
\end{align}
where $m,n,\gamma,\epsilon$ are free parameters. The physical significance of these parameters is best understood in limiting cases of this family of solutions, where, in appropriate combinations, they can be associated to mass, NUT charge, angular momentum and acceleration.

In \cite{Plebanski:1976gy}, Plebanski and Demianski also point out the remarkable fact that the family of solutions they found admits a double-Kerr-Schild form, if we allow for a complex extension of the coordinates, that is, for a complex manifold. Our approach is to allow for this complexification and to employ the double-Kerr-Schild coordinates in order to identify a single-copy gauge field. We can thereafter interpret the gauge field by returning to the real section. The reason why double-Kerr-Schild coordinates are helpful is that they provide us with a map between the curved spacetime and the flat spacetime on which the gauge field lives. Consider the coordinate transformation 
\be
\tau = u + \int\frac{q^2dq}{ Q(q)}+i\int \frac{p^2dp}{ P(p) } \,, \qquad   
\sigma = v - \int\frac{dq}{ Q(q)}+i\int \frac{dp}{ P(p) }\,.
\ee
The metric is now given by
\be
\label{PDdoubleKS1}
ds^2 =  \frac{1}{(1-pq)^2} \!\! \left[2 i (du+q^2 dv) dp -2(du-p^2 dv) dq   + \frac{P(p)}{(p^2+q^2)} (du+q^2 dv)^2 - \frac{Q(q)}{(p^2+q^2)}(du-p^2 dv)^2 \right]  
\ee
so that it becomes linear in the free parameters $m,n,\gamma,\epsilon$. We have therefore the double-Kerr-Schild form
\be
\label{PDdoubleKS}
ds^2 = ds^2_{(0)} + \phi_{K} \, K^2 + \phi_{L} \, L^2 \,,
\ee
where $K$ and $L$ are null, geodesic and mutually orthogonal covectors,
\be
K=du+q^2 dv \,, \qquad L=du-p^2 dv \,,
\ee
and the metric $ds^2_{(0)}$ is flat. 

There is an ambiguity in the splitting between $ds^2_{(0)}$ and the Kerr-Schild functions $\phi_{K}$ and $\phi_{L}$, because any metric for which both $m$ and $n$ vanish is flat. We will see this later in the expression for the Weyl spinor. Such parameters $m$ and $n$ are sometimes called {\it dynamical} parameters, whereas the parameters that do not generate curvature are called {\it kinematical} parameters, in our case $\gamma$ and $\epsilon$. We choose to put the kinematical parameters into the flat metric $ds^2_{(0)}$, and some of our later results will depend crucially on this choice. We have therefore
\be
\label{PDdoubleKSflat}
ds^2_{(0)} =  \frac{1}{(1-pq)^2} \left[2 (i K dp -L dq)   + \frac{\gamma (1-p^4) - \epsilon p^2}{(p^2+q^2)} \,K^2 - \frac{ \gamma (1-q^4) + \epsilon q^2}{(p^2+q^2)}\,L^2 \right]  \,,
\ee
and
\be
\phi_{K} = \frac{2 n p +2 m p^3}{(1-pq)^2(p^2+q^2)} \,, \qquad 
\phi_{L} = \frac{2m q + 2 n q^3}{(1-pq)^2(p^2+q^2)} \,.
\ee

While a Kerr-Schild metric has a linear Ricci tensor $R^\mu_{\;\;\,\nu}$, as described in section~\ref{sec:reviewKS}, this property is not guaranteed for a double-Kerr-Schild metric. It turns out that this property holds nonetheless for the solution at hand, in the following sense. Let us introduce parameters $c_K$ and $c_L$ multiplying $\phi_K$ and $\phi_L$, respectively, in \eqref{PDdoubleKS}. Then $R^\mu_{\;\;\,\nu}$ does not vanish if $c_K$ and $c_L$ are distinct, but is given by the difference between the Ricci tensors of the single-Kerr-Schild metrics with either the $K$ term or the $L$ term, that is,
$$
R^\mu_{\;\;\,\nu} [c_K\phi_K K^2 + c_L\phi_L L^2] = R^\mu_{\;\;\,\nu} [c_K\phi_K K^2 ] + R^\mu_{\;\;\,\nu} [c_L\phi_L  L^2] = (c_K-c_L)\, R^\mu_{\;\;\,\nu} [\phi_K K^2 ] \,.
$$

We can now try to apply the (double) Kerr-Schild double copy, as was done in \cite{Luna:2015paa} for the Taub-NUT solution. That is, we seek functions $a_{K}$ and $a_{L}$ such that the gauge field
\be
\label{gaugePDdKS}
A= a_{K} \, K + a_{L} \, L 
\ee
satisfies the Maxwell equations in the flat spacetime $ds^2_{(0)}$. This is not the case if we identify $a_{K}$ and $a_{L}$ with $\phi_{K}$ and $\phi_{L}$, which would be the naive guess following  section~\ref{sec:reviewKS}; indeed, time independence played a crucial role there. While there is in fact a range of choices of $a_{K}$ and $a_{L}$ such that $A$ satisfies the Maxwell equations, we will see that the Weyl double copy picks up a particular solution.

We take the following vierbein for the curved metric \eqref{PDdoubleKS1} with coordinates $(u,v,p,q)$:
\be
e^a_{\phantom{a}\mu} = \frac1{\sqrt{2}(1-pq)} \left( \begin{array}{cccc}
1+\mathcal{Q} & -p^2(1+\mathcal{Q}) & 0 & 1 \\
1-\mathcal{Q} & -p^2(1-\mathcal{Q}) & 0 & -1 \\
1+\mathcal{P} & q^2(1+\mathcal{P}) & i & 0 \\
i(1-\mathcal{P}) & iq^2(1-\mathcal{P}) & 1 & 0
\end{array} \right),
\quad 2\,\mathcal{Q}=\frac{Q(q)}{p^2+q^2}\,,
\quad 2\,\mathcal{P}=\frac{P(p)}{p^2+q^2}\,.
\ee
The Weyl spinor is
\be
C_{ABCD} \,\xi^A\xi^B\xi^C\xi^D = -\frac{3i}{2}\, \frac{(m+in)(1-pq)^3\left((\xi^1)^2-(\xi^2)^2\right)^2}{(p-iq)^3}.
\ee
It is clear now that $m$ and $n$ are dynamical parameters, while $\gamma$ and $\epsilon$ are kinematical parameters.

For the gauge field in flat spacetime ($m=n=0$), we seek a field strength spinor of the analogous form
\be
\label{eq:fABtypeD}
f_{AB}\, \xi^A\xi^B = -\frac{1}{2}\, \frac{({\mathfrak m}+i{\mathfrak n})(1-pq)^2\left((\xi^1)^2-(\xi^2)^2\right)}{(p-iq)^2},
\ee
with free parameters ${\mathfrak m}$ and ${\mathfrak n}$. The reasoning for the single power of $\left((\xi^1)^2-(\xi^2)^2\right)$ should be clear. As for the power of $(1-pq)/(p-i q)$, 
this choice precisely matches equation \eqref{eq:penrosewalker}; in particular,
\be
\chi_{AB}\, \xi^A\xi^B =  \frac{(p-iq)\left((\xi^1)^2-(\xi^2)^2\right)}{(1-pq)}
\ee
up to a constant factor.

The numerical pre-factor in \eqref{eq:fABtypeD} was chosen so that the gauge field can be written as
\be
\label{singlecopyPD}
A= \frac{{\mathfrak n}\,p}{p^2+q^2} \, K + \frac{{\mathfrak m}\,q}{p^2+q^2} \, L \,.
\ee
It takes the form anticipated in \eqref{gaugePDdKS}, and it satisfies the Maxwell equations. The Weyl double copy \eqref{eq:Wdoublecopy} also identifies
\be
\label{zerocopyPD}
S= \frac{i}{6}\, \frac{({\mathfrak m}+i{\mathfrak n})^2(1-pq)}{(m+in)(p-iq)} \,,
\ee
which satisfies the wave equation in the flat background.

\subsection{Scaling limits}
\label{sec:scaling}

Having accomplished our goal of obtaining the single copy for the general family of vacuum type D metrics, let us now focus on particular cases for illustration. These cases follow from certain scaling limits of the coordinates and parameters.

\subsubsection{Kerr-Taub-NUT metric}

The first particular case that we will consider in detail is the Kerr-Taub-NUT solution \cite{Carter:1968rr,Plebanski:1975xfb}, which includes mass, angular momentum and NUT charge. We see it here as a limit of the general solution \eqref{PDdoubleKS1}. Consider the coordinate scaling
\be
u\to \ell u\,, \quad v\to \ell^{3}v\,, \quad  p\to \ell^{-1}p\,, \quad q\to \ell^{-1}q\,, 
\ee
and the parameter scaling
\be
m\to \ell^{-3} m\,, \quad n\to \ell^{-3}n\,, \quad  \epsilon\to \ell^{-2}\epsilon\,, \quad \gamma\to \ell^{-4}\gamma\,, 
\ee
and now take $\ell \to \infty$. The result is
\begin{align}
\label{taubnutmetriclimit}
ds^2 = \; & 2 (i K dp -L dq)
+ \gamma\,  (2 du + dv (q^2-p^2)) dv 
- \epsilon\, (du^2+p^2q^2dv^2)  \nonumber \\
& + \frac{2n p}{p^2+q^2} \,K^2+ \frac{2m q}{p^2+q^2} \,L^2 \,,
\end{align}
where the first line denotes $ds^2_{(0)}$. This solution has lost a free parameter associated to acceleration: if $\epsilon \neq 0$, we can choose $\epsilon$ to take only the values $1$ or $-1$ by a rescaling of the form
\begin{align}
& u\to |\epsilon|^{-1/2} u\,, \quad v\to |\epsilon|^{-3/2} v\,, \quad  p\to |\epsilon|^{1/2} p\,, \quad q\to |\epsilon|^{1/2} q\,, \nonumber \\
& m\to |\epsilon|^{3/2} m\,, \quad n\to |\epsilon|^{3/2} n\,, \quad \gamma\to |\epsilon|^{2} \gamma\,.
\end{align}

The single copy is now obtained by applying the same scaling limit to \eqref{singlecopyPD}, together with $({\mathfrak m},{\mathfrak n})\to \ell^{-3} ({\mathfrak m},{\mathfrak n})$, and by taking the leading coefficient $\mathcal{O}(\ell^{-1})$,
\be
\label{taubnutgaugelimit}
A= \frac{{\mathfrak n}\, p}{p^2+q^2} \,K+ \frac{{\mathfrak m}\, q}{p^2+q^2} \,L \,.
\ee
Finally, we do the same for \eqref{zerocopyPD} with $\mathcal{O}(\ell^{-2})$,
\be
S= \frac{i}{6}\, \frac{({\mathfrak m}+i{\mathfrak n})^2}{(m+in)(p-iq)} \,,
\ee

These results reproduce the double-copy relation between the Taub-NUT metric and the dyonic gauge field (electric charge and magnetic monopole charge) identified in \cite{Luna:2015paa}. It appears here as a particular case of the larger Plebanski-Demianski family\footnote{The double-Kerr-Schild scalar identified in \cite{Luna:2015paa} is a linear combination of the real and imaginary parts of $S$.}.

\subsubsection{C-metric}

The second particular case of our vacuum type D double copy that we want to focus on is the C-metric, which can be interpreted as describing a pair of uniformly accelerated black holes; see e.g. \cite{Griffiths:2006tk} for a detailed discussion of this solution. The C-metric is studied here for the first time in the context of the double copy. It can also be obtained from a scaling limit. After performing the coordinate change
\be
p=x\, \qquad q=-\frac1{y}\,,
\ee
the scaling limit is $\ell\to\infty$ with
\be
\label{scalingCmetric}
(u,v,x,y)\to \ell^{-1} (u,v,x,y)\,, \quad m\to \ell^3\, m\,, \quad n\to \ell\, n\,, \quad \gamma\to \gamma\,, \quad  \epsilon\to \ell^{2}\epsilon\,.
\ee
Applied to the metric \eqref{PDdoubleKS1}, the result is
\be
\label{Cmetric}
ds^2 = \frac{2}{(x+y)^2}(-du\,dy+idv\,dx) + \frac{G(x)}{(x+y)^2}\, dv^2 - \frac{F(y)}{(x+y)^2} \, du^2 \,,
\ee
where
\begin{align}
&  G(x) =  \gamma +2 n x- \epsilon x^2 +2 m x^3 \,, \nonumber \\
&  F(y) =  -\gamma+ 2n y + \epsilon y^2 +2 m y^3 \,.
\end{align}
The Weyl spinor
\be
C_{ABCD} \,\xi^A\xi^B\xi^C\xi^D = \frac{3}{2}\, m\,(x+y)^3\left((\xi^1)^2-(\xi^2)^2\right)^2
\ee
shows that $m$ is now the only dynamical parameter.

In order to obtain the best-known form of the C-metric \cite{Kinnersley:1970zw}, we apply the coordinate transformation
\be
\label{coordut}
u =\text{t} - \int\frac{dy}{ F(y)} \,, \qquad   
v = \phi - i \int \frac{dx}{ G(x) }\,,
\ee
and introduce a parameter $A$ via the rescaling 
\be
(\text{t},\phi)\to A^{-2} (\text{t},\phi)\,, \quad (m,n,\gamma,\epsilon)\to A^2 (m,n,\gamma,\epsilon)\,,
\ee
which leads to
\be
\label{usualCmetric}
ds^2= \frac{1}{A^2(x+y)^2} \left[
-F(y) \,d\text{t}^2 + \frac{dy^2}{F(y)} + \frac{dx^2}{G(x)} + G(x)\, d\phi^2
\right]\,.
\ee
Moreover, we restrict our parameters to match
\begin{align}
&  G(x) = 1-x^2-2A \,\mathfrak{M} \,x^3 \,, \nonumber \\ 
&  F(y) = -1+y^2-2A\,  \mathfrak{M}\, y^3 \,, 
\end{align}
so that
\be
\label{Cparameters}
m=-A \,\mathfrak{M}\,, \quad n=0\,, \quad \gamma=\epsilon=1 \,. 
\ee

The physical interpretation of the C-metric is clearer in a different coordinate system, where we can contrast it with the Schwarzschild solution. Firstly, there is a linear change of coordinates \cite{Hong:2003gx},
\be
(\text{t},x,y,\phi) = \left(\frac{c_0}{B}\,\tilde {\text{t}}\,,\,Bc_0(\tilde x-c_1)\,,\,Bc_0(\tilde y+c_1)\,,\,\frac{c_0}{B}\,\tilde\phi\right)  \,,
\ee
where $B,c_0,c_1$ are constants, and a change of parameters
\be
A = B^{-1}\,\alpha  \,,  \qquad \mathfrak{M} = c_0^{-3}\,M\,,
\ee
that put the metric in the form analogous to \eqref{usualCmetric} but with\footnote{The transformation is trivial ($B=c_0=1,c_1=0$) if $M=0$, i.e., in the flat case. The explicit constants can be expressed as
\be
B^2 = \frac{1-36\,\alpha^2 M^2+(1+12\,\alpha^2 M^2)^{3/2}}{2\,(1-4\,\alpha^2 M^2)^2} \,,
\qquad c_0^4=1+12\,\alpha^2 M^2 \,, \qquad c_1 =\frac{\sqrt{1+12\,\alpha^2 M^2}-1}{6\,\alpha M}\,.
\ee
 }
\begin{align}
&  \tilde G(\tilde x) = (1-\tilde x^2)(1+2\alpha  M \tilde x) \,, 
\nonumber \\
&  \tilde F(\tilde y) = -(1-\tilde y^2)(1-2\alpha  M \tilde y) \,.
\end{align} 
Secondly, we introduce
\be
\label{coordxyt}
\tilde x=-\cos\theta \,, \quad \tilde y=\frac1{\alpha r} \,, \quad \tilde {\text{t}}=\alpha t \,,
\ee
to obtain
\be
ds^2 = \frac1{(1-\alpha r \cos\theta)^2} \left[
-f(r) dt^2 + \frac{dr^2}{f(r)} + r^2 \left( \frac{d\theta^2}{g(\theta)} + g(\theta)\sin^2\theta d\phi^2  \right)
\right]\,,
\ee
with
\be
f(r)=\left(1-\frac{2M}{r}\right)(1-\alpha^2r^2)\,, \qquad g(\theta)=1-2\alpha M \cos\theta \,.
\ee
For $\alpha=0$, we recover the Schwarzschild black hole with mass $M$. In general, the solution has a black hole horizon at $r=2M$ and an acceleration horizon at $r=1/\alpha$. It can be interpreted as a black hole with uniform acceleration $\alpha$, where this acceleration is provided by a cosmic string (notice that it is not possible to make both half-axes $\theta=0$ and $\theta=\pi$ regular). In fact, a global analysis of this solution indicates that there is a pair of causally-disconnected black holes uniformly accelerating in opposite directions. There is either a cosmic string between them pushing them apart, or two semi-infinite cosmic strings pulling them apart. See \cite{Griffiths:2006tk} for more details, including the Penrose diagrams describing the global structure of the solution.

Getting back to our double-copy story, let us consider the single copy gauge field. It is obtained by applying the scaling limit \eqref{scalingCmetric} to the solution \eqref{singlecopyPD}, together with ${\mathfrak m}\to \ell^{3} {\mathfrak m}, \, {\mathfrak n}\to \ell {\mathfrak n}$, and by taking the leading coefficient $\mathcal{O}(\ell)$, yielding
\be
\label{singlecopyCmetric}
A= - {\mathfrak m} \,y\, du  \,.
\ee
Notice that the parameter ${\mathfrak n}$ did not survive the limit, where we chose the scaling of ${\mathfrak m},{\mathfrak n}$ to be the same as that of $m,n$, respectively. For the field strength spinor, we have
\be
f_{AB}\, \xi^A\xi^B = \frac{1}{2}\, {\mathfrak m}\,(x+y)^2\,\left((\xi^1)^2-(\xi^2)^2\right) \,.
\ee
Finally, we do the same for \eqref{zerocopyPD} with $\mathcal{O}(\ell^{2})$,
\be
S= \frac{{\mathfrak m}^2}{6 m}\, (x+y) \,.
\ee

\begin{figure}
\def\axisdefaultwidth{200pt}
\def\axisdefaultheight{200pt}
\centering
\begin{tikzpicture}
	\begin{axis}[xmin=-2.35,xmax=2.35,ymin=-2.35,ymax=2.35]
		\addplot [thick,domain=-1.5:1.5] ({cosh(x)}, {sinh(x)});
		\addplot [thick,domain=-1.5:1.5] ({-cosh(x)}, {sinh(x)});
		\addplot[dashed] expression {x};
		\addplot[dashed] expression {-x};
	\end{axis}
	\node [rotate=46] at (0.2, 0.8) {$r = 0$};
	\node [rotate=45] at (4.8,5.2) {$r= 1/\alpha$};
\end{tikzpicture}
\caption{The curved lines ($r=0$) represent two causally disconnected uniformly accelerated particles. The diagonal lines  ($r=1/\alpha$) represent the acceleration horizons, which split the plot into four quadrants.}
\label{fig:accelerated}
\end{figure}
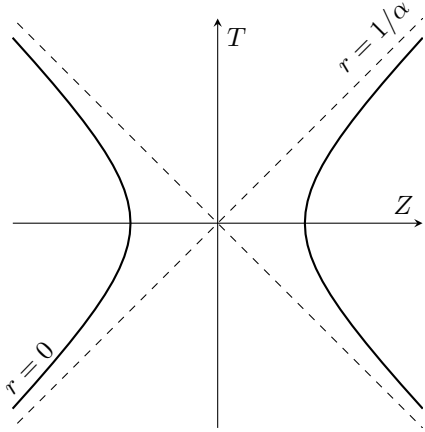

Let us interpret the gauge field \eqref{singlecopyCmetric}, which lives on flat spacetime. Changing coordinates by analogy with \eqref{coordut}, and recalling the parameters \eqref{Cparameters} in the flat case ($M=0$), we get
\be
\label{cmetricgaugefield}
A= - {\mathfrak m} \,y\, du = - {\mathfrak m}  \left( y \,d\text{t} - \frac{y\, dy}{-1+y^2} \right) \cong - {\mathfrak m}  \,y \,d\text{t} = Q \,\frac{dt}{r} \,,
\ee
where we used  $\cong$ to express equivalence up to gauge freedom, and suggestively relabelled the final free parameter. In the final step, we also used the analogue of \eqref{coordxyt}. (Notice that, while we reused $\text{t}$ and $t$ in analogy to the curved case in a slight abuse of notation, these two coordinates do not map trivially from the C-metric to Minkowski spacetime, unlike the double-Kerr-Schild coordinates $u$ and $y$.) We are left with a solution resembling the Coulomb potential. However, we are using `accelerated' coordinates, with the Minkowski metric given by
\be
ds^2_{(0)} = \frac1{(1-\alpha r \cos\theta)^2} \left[
-(1-\alpha^2r^2) dt^2 + \frac{dr^2}{1-\alpha^2r^2} + r^2 \left( d\theta^2 +\sin^2\theta d\phi^2  \right)
\right]\,.
\ee
If we choose the acceleration to be along the $Z$ axis, we can write the Minkowski metric as
\be
ds^2_{(0)} = -dT^2+dX^2+dY^2+dZ^2\,,
\ee
and the gauge field as\,,
\be
\label{cgaugefield}
A=\frac{-T^2+X^2+Y^2+Z^2+\alpha^{-2}}{ \sqrt{\left(-T^2+X^2+Y^2+Z^2+\alpha^{-2}\right)^2 -4\alpha^{-2}\left(-T^2+Z^2\right) }}\frac{T\, dZ-Z\, dT}{T^2-Z^2}\,,
\ee
with
\be
\label{coordTZXY}
(T',Z') =\frac1{\alpha}\,\frac{\sqrt{|1-\alpha^2r^2|}}{1-\alpha r \cos\theta} \left(\sinh \alpha t,\cosh \alpha t\right) \,, \qquad (X,Y) =\frac{r\sin{\theta}}{1-\alpha r \cos\theta} \left(\cos\phi,\sin\phi\right) \,,
\ee
where, according to figure~\ref{fig:accelerated}, $(T,Z)=(T',Z')$ in the right quadrant, $(T,Z)=-(T',Z')$ in the left quadrant, $(T,Z)=(Z',T')$ in the top quadrant, and $(T,Z)=-(Z',T')$ in the bottom quadrant. Notice that $r$ is not a global coordinate, and that $r=0$ represents two worldlines. It is clear now that we are dealing with a pair of causally disconnected charges uniformly accelerating in opposite directions with acceleration $\alpha$, and that the solution is just the corresponding Li\'enard-Wiechert potential.

For complete clarity, we can also match our gauge field directly to the best-known formula for the Li\'enard-Wiechert potential. To do so, we use flat spacetime coordinates centred on one of the charges. We take the one in the right quadrant of figure~\ref{fig:accelerated}. Let us start with standard Cartesian coordinates $X^\mu=(T,X,Y,Z)$ and consider the worldline $X^\mu=X_p^\mu(U)$, where $U$ is the proper time, such that $\dot{X_p}{}^2=-1$. Now, we change to a null coordinate system $(U,R,w,\wb)$ associated to the family of light-cones emanating from the worldline,
\be
\label{coordNewman}
X^\mu=X_p^\mu(U)+\frac{R\,\ell^\mu(w,\wb)}{\chi(U,w,\wb)} \,,
\ee
where $R$ is the retarded distance, $\chi = -\ell \cdot \dot{X_p}$, and
\be
l^\mu = \left(  1, \frac{w+\wb}{1+w\wb} , \frac{i(\wb-w)}{1+w\wb},\frac{-1+w\wb}{1+w\wb}\right)
\ee
is a null vector giving the angular direction. The Minkowski metric then reads
\be
\label{eq:adapted}
ds^2_{(0)} = -\left( 1-2\,R\,\frac{\dot{\chi}}{\chi} \right) \,dU^2 -2 dU\, dR +\frac{4\,R^2}{(1+w\wb)^2\,\chi^2} dw d\wb \,.
\ee
The typical expression for the Li\'enard-Wiechert potential is (with our `mostly-plus' metric signature)
\be
A^{(\text{LW})} = -Q\, \frac{\dot{X_p}^\mu(U)}{R} \,dX_\mu \,.
\ee
For uniformly accelerated motion in the $Z$ direction, we have
\be
\dot{X_p}^\mu(U)= ( \cosh\alpha U,0,0,\sinh \alpha U),
\ee
and therefore
\be
A^{(\text{LW})} =Q\, \frac{ \cosh\alpha U\, dT - \sinh \alpha U\, dZ}{R}  \,.
\ee
Inverting the coordinate change \eqref{coordNewman}, it is possible to compare this to  \eqref{cgaugefield} and to verify that
\be
A \cong A^{(\text{LW})}  \,,
\ee
i.e., that our single-copy gauge field is equivalent to the Li\'enard-Wiechert potential up to a gauge transformation.

In conclusion, we found that the C-metric, which represents a pair of uniformly accelerated black holes, is the double copy of the Li\'enard-Wiechert potential for a pair of uniformly accelerated charges. This is the natural expectation. Our investigations were motivated by an analogy pointed out by Newman \cite{Newman:1973} between the Robinson-Trautman family, which includes the C-metric, and the Li\'enard-Wiechert potential. Here, we have shown how to turn this analogy into an exact map between solutions.

\section{A complex example: the Eguchi-Hanson instanton} \label{sec:EH}

Self-dual solutions provide a particularly simple setting in which to understand the double copy \cite{Monteiro:2011pc}. In this section, we will focus on the Eguchi-Hanson gravitational instanton \cite{Eguchi:1978xp,Eguchi:1978gw}, motivated by the recent work \cite{Berman:2018hwd} on the double-copy interpretation of that solution. We will see that the Eguchi-Hanson solution can be understood in two distinct ways through the double copy. One follows the approach that we have taken so far in this paper, that of a `pure' Weyl double copy associated with a unique gauge field solution. The other follows the approach taken in ref.~\cite{Berman:2018hwd}, which we reinterpret here as a `mixed' Weyl double copy, involving two gauge field solutions related by a coordinate exchange.

\subsection{Symmetric Weyl double copy}

It can be shown \cite{Tod,Berman:2018hwd} that the Eguchi-Hanson spacetime can be written in the Kerr-Schild form
\be
\label{EHdoubleKS}
ds^2= 2dudv-2dXdY+\frac{\lambda}{(uv-XY)^3}(vdu-XdY)^2.
\ee
The flat background metric $ds^2_{(0)}$ is simply
\be
\label{EHdoubleKSflat}
ds^2_{(0)} = 2dudv-2dXdY,
\ee
while the Kerr-Schild covector (using coordinates $(u,v,X,Y)$) and the scalar field correspond to
\begin{eqnarray}
k_\mu=\frac{1}{(uv-XY)}(v,0,0,-X),\qquad \phi=\frac{\lambda}{(uv-XY)}.
\end{eqnarray}
The vierbein can be given by
\begin{align}
\nonumber e^0_{\ \mu}&=\frac{i}{\sqrt{2}}\left((1,1,0,0)+\frac{v}{2}\zeta\right), \qquad
e^1_{\ \mu}=\frac{i}{\sqrt{2}}\left((1,-1,0,0)-\frac{v}{2}\zeta\right), \\
 e^2_{\ \mu}&=\frac{i}{\sqrt{2}}\left((0,0,1,1)+\frac{X}{2}\zeta\right),\qquad
 e^3_{\ \mu}=\frac{1}{\sqrt{2}}\left((0,0,-1,1)-\frac{X}{2}\zeta\right),
 \label{tetradfullEH}
\end{align}
where
\begin{eqnarray}
\zeta=\frac{\lambda}{(uv-XY)^3}(v,0,0,-X).
\end{eqnarray}
Using this vierbein, we can compute the Weyl spinor, and the result is
\begin{align}
C_{ABCD}=0,
\end{align}
which follows from the self-duality of the solution -- it is not a real solution in Lorentzian signature.
On the other hand, we have the non-trivial self-dual component\footnote{In this section, we use $\tilde{C}_{\dA\dB\dC\dD} $ instead of $\bar{C}_{\dA\dB\dC\dD}$ to emphasise that it is not the complex conjugate of  $C_{ABCD}$, since the solution is complex.}
\begin{eqnarray}
\tilde{C}_{\dA\dB\dC\dD} = \frac{1}{4} W_{\mu\nu\rho\lambda}\,\tilde{\sigma}^{\mu\nu}_{\dA\dB}\,\tilde{\sigma}^{\rho\lambda}_{\dC\dD},
\end{eqnarray}
where $\tilde{\sigma}^{\mu\nu}_{\dA\dB}= \sigma^{[\mu}_{A\dA} \,{\tilde{\sigma}}^{\nu]\; \dC A}\, \eps_{\dC\dB}$\,. Explicitly,
\begin{eqnarray}
\tilde{C}_{\dA\dB\dC\dD}\xi^{\dA}\xi^{\dB}\xi^{\dC}\xi^{\dD}=\frac{-3\lambda(-iX(\xi_1-\xi_2)+v(\xi_1+\xi_2))^2(u(\xi_1-\xi_2)+iY(\xi_1+\xi_2))^2}{2(uv-XY)^5}.
\label{WeylEH}
\end{eqnarray}

It turns out that the naive gauge field obtained from the Kerr-Schild metric,
\be
\label{singlecopydoubleKS}
A_\mu=\phi k_\mu=\frac{\lambda}{(uv-XY)^2}(v,0,0,-X)\,,
\ee
satisfies the Maxwell equations and leads to the correct Weyl double copy. We can compute the field strength spinor and anti-spinor using the flat-metric version of  the vierbein \eqref{tetradfullEH} (i.e., no $\zeta$ term). This leads to
\begin{eqnarray}
f_{AB}= 0\,,
\end{eqnarray}
which is due to self-duality,
and to
\begin{eqnarray}
\tilde{f}_{\dA\dB} = \frac{1}{2} F_{\mu\nu}\, \tilde{\sigma}^{\mu\nu}_{\dA\dB}\
\end{eqnarray}
such that
\begin{eqnarray}
\tilde{f}_{\dA\dB}\, \xi^{\dA}\xi^{\dB} =-\lambda\frac{(-iX(\xi_1-\xi_2)+v(\xi_1+\xi_2))(u(\xi_1-\xi_2)+iY(\xi_1+\xi_2))}{(uv-XY)^3}\,.
\label{FSEH}
\end{eqnarray}
From (\ref{WeylEH}) and (\ref{FSEH}), it is clear that the Weyl double-copy relation is satisfied once we identify $\tilde{S}$ with $\phi$ as the scalar field, up to a numerical factor; notice that $\tilde{S}$ is real in Lorentzian spacetime. The scalar field $\phi$ clearly satisfies the wave equation.

\subsection{Mixed Weyl double copy}

There is another way in which we can imagine understanding the double copy. Consider a general type D Weyl anti-spinor, 
\be
\tilde{C}_{\dA\dB\dC\dD} =  \frac1{\tilde{S}} \;\; \mathbf{\tilde a}_{(\dA} \,\mathbf{\tilde b}_{\dB}  \,\mathbf{\tilde a}_{\dC} \,\mathbf{\tilde b}_{\dD)}.
\ee
In the preceding approach, we took\, $\tilde{f}_{\dA\dB} = \mathbf{\tilde a}_{(\dA} \,\mathbf{\tilde b}_{\dB)}$\,, but we could also consider
\be
\label{eq:mixedWeyl}
\tilde{C}_{\dA\dB\dC\dD} = \frac1{\tilde{S}}\, {\tilde m}_{(\dA\dB} { \tilde m'}_{\dC\dD)}\,, 
\ee
with
\be
\tilde m_{\dA\dB} =\mathbf{\tilde a}_{\dA} \,\mathbf{\tilde a}_{\dB} \,, \qquad {\tilde m'}_{\dA\dB}=\mathbf{\tilde b}_{\dA} \,\mathbf{\tilde b}_{\dB}\,.
\ee
In this way, the field strengths $\tilde m_{\dA\dB}$ and ${\tilde m'}_{\dA\dB}$ have a single principal null direction of multiplicity 2. In the case of the Eguchi-Hanson instanton, we have
\begin{eqnarray}
\tilde m_{\dA\dB}\, \xi^{\dA}\xi^{\dB} =-\lambda\,\frac{i(-iX(\xi_1-\xi_2)+v(\xi_1+\xi_2))^2}{2(uv-XY)^3}\,.
\label{mspinor}
\end{eqnarray}
and 
\begin{eqnarray}
{\tilde m'}_{\dA\dB}\, \xi^{\dA}\xi^{\dB} =-\lambda\,\frac{i(u(\xi_1-\xi_2)+iY(\xi_1+\xi_2))^2}{2(uv-XY)^3}\,.
\label{mpspinor}
\end{eqnarray}

In ref.~\cite{Berman:2018hwd}, the gauge field $A^{(\tilde m)}_{\mu}$ corresponding to $\tilde m_{\dA\dB}$ (with $m_{AB}$=0), which satisfies the Maxwell equations, was proposed to be the single copy of the Eguchi-Hanson instanton. The reasoning is that, in terms of the differential operator 
\be
\hat{k}_\mu = (\partial_X,0,0,\partial_v)\,,
\ee
 the Eguchi-Hanson metric \eqref{EHdoubleKS} and the gauge field are
\be
g_{\mu\nu} ={g_{(0)}}_{\mu\nu} + \hat{k}_\mu  \hat{k}_\nu \Theta \,, \qquad 
A^{(\tilde m)}_{\mu} =\hat{k}_\mu \Theta \,, \qquad \Theta=\frac{X^2}{2\,u^2} \frac{\lambda}{(uv-XY)} = \frac{X^2}{2\,u^2} \,\phi \,.
\ee
The explicit form of the gauge field is
\be
\label{singlecopydoubleKS}
A^{(\tilde m)}_{\mu} =\frac{\lambda\, X}{2u(uv-XY)^2}\left(\frac{2uv-XY}{u},0,0,-X\right)\,.
\ee
If $\Theta$ is identified instead with $\phi \,v^2/(2Y^2)$, then we obtain the same metric and (up to a gauge transformation) the same gauge field.

Our construction makes it clear that there is an analogous gauge field solution $A^{(\tilde m')}_{\mu}$ corresponding to $\tilde{m}'_{\dA\dB}$ (with ${m}'_{AB}$=0). In terms of the differential operator
\be
\hat{k}'_\mu = (0,\partial_Y,\partial_u,0)\,,
\ee
we obtain a metric and a gauge field,
\be
g'_{\mu\nu} ={g_{(0)}}_{\mu\nu} + \hat{k}'_\mu  \hat{k}'_\nu \Theta' \,, \qquad 
A^{(\tilde m')}_{\mu} =\hat{k}'_\mu \Theta' \,, \qquad \Theta'=\frac{Y^2}{2\,v^2} \frac{\lambda}{(uv-XY)} = \frac{Y^2}{2\,v^2}\, \phi \,.
\ee
The metric is the Eguchi-Hanson metric \eqref{EHdoubleKS} with the coordinate exchange $u\leftrightarrow v$, $X\leftrightarrow Y$, which, given the corresponding change in the vierbein, still leads precisely to \eqref{WeylEH}. 

The relation \eqref{eq:mixedWeyl} expresses a {\it mixed} Weyl double copy, with field strength anti-spinors $\tilde{m}_{\dA\dB}$ and $\tilde{m}'_{\dA\dB}$. The {\it pure} Weyl double copy associated with $\tilde{m}_{\dA\dB}$ is a type N solution with Weyl anti-spinor
\be
\tilde C^{(\tilde m)}_{\dA\dB\dC\dD} = \frac1{\tilde{S}}\, {\tilde m}_{(\dA\dB} { \tilde m}_{\dC\dD)} 
= \frac1{\tilde{S}}\, \mathbf{\tilde a}_{(\dA} \,\mathbf{\tilde a}_{\dB}  \,\mathbf{\tilde a}_{\dC} \,\mathbf{\tilde a}_{\dD)}\,.
\ee
Explicitly,
\begin{eqnarray}
\tilde C^{(\tilde m)}_{\dA\dB\dC\dD}\xi^{\dA}\xi^{\dB}\xi^{\dC}\xi^{\dD}=\frac{3\lambda(-iX(\xi_1-\xi_2)+v(\xi_1+\xi_2))^4}{2(uv-XY)^5}.
\end{eqnarray}
Likewise, the pure Weyl double copy associated with $\tilde{m}'_{\dA\dB}$ is a type N solution with Weyl anti-spinor
\begin{eqnarray}
\tilde C^{(\tilde m')}_{\dA\dB\dC\dD}\xi^{\dA}\xi^{\dB}\xi^{\dC}\xi^{\dD}=\frac{3\lambda(u(\xi_1-\xi_2)+iY(\xi_1+\xi_2))^4}{2(uv-XY)^5}.
\end{eqnarray}
The algebraic relation of these type N solutions to \eqref{WeylEH} is clear. Each of the two type N solutions possesses a single principal null direction (of multiplicity 4), and these correspond precisely to the pair of principal null directions (of multiplicity 2) of the Eguchi-Hanson instanton.

\section{Discussion} \label{sec:discussion}

In this paper, we have described how the double copy for a certain class of exact solutions to the field equations can be described in terms of the curvatures, rather than the fields -- a prescription we called the Weyl double copy. This extends the range of examples of the classical double copy to a larger family of solutions, including general vacuum type D spacetimes. A notable example is the C-metric, which we worked out in detail. On the one hand, the Weyl double copy provides a unique correspondence in cases where the Kerr-Schild double copy is ambiguous, as we discussed for pp-waves. On the other hand, it also allows for a wider range of double-copy connections between gravity and gauge theory solutions, which we explored in the example of the Eguchi-Hanson instanton.

We leave several questions open. An obvious one within type D is how to introduce a cosmological constant, and electric and magnetic charges. A satisfactory understanding of these cases must include a prescription for the Ricci tensor, and not just the Weyl tensor. The spinorial approach provides a natural path forward here. There are other matter systems of interest, however. The double copy should naturally incorporate a scalar dilaton field and a two-form field. An important result in this direction was presented recently in \cite{Lee:2018gxc}, where an extended Kerr-Schild ansatz including those fields was proposed in the context of double field theory, allowing for a Kerr-Schild-type double-copy prescription. It would be interesting to know whether the Weyl double copy can be similarly extended. A sophisticated extension of the Weyl double copy would include non-type D spacetimes. Another natural direction is the application to higher-dimensional solutions, which motivated the recent paper \cite{Monteiro:2018xev} extending to higher dimensions the spinorial approach to general relativity.

A fundamental question is why the classical double copy of exact solutions, either in the Kerr-Schild prescription or in the Weyl prescription, is even possible. The double copy for scattering amplitudes is formulated in momentum space, and therefore one expects that, generically, a coordinate space prescription is non-local. A sensible proposal in the case of linearised fields was presented in \cite{Anastasiou:2014qba}, where a convolution procedure is employed. The crucial point is the role of the scalar field in relating the gauge field and the gravitational field. Presumably, for the algebraically special cases considered so far in the double copy of exact solutions, the non-locality effectively disappears. Progress in understanding this would give an important clue into a (possibly quite sophisticated) formulation of the double copy of generic solutions. 

To conclude, we hope that our paper will bring the classical double copy to the attention of the general relativity community, whose accumulated body of work could have a rapid impact on the topic.


\acknowledgments

We would like to thank Tim Adamo, Graham Brown and Lionel Mason for discussions. We also thank David Peinador Veiga and Brian Kent for pointing out an incorrect statement in section 3 in earlier versions of the paper. AL is supported in part by the Department of Energy under Award Number DESC000993, and thanks Chris White and Erick Chac\'on for discussions and collaboration on related topics. RM is supported by a Royal Society University Research Fellowship. IN is supported by STFC studentship ST/N504051/1. DOC is an IPPP associate, and thanks the IPPP for on-going support as well as for hospitality during this work. He is supported in part by the Marie Curie FP7 grant 631370 and by the STFC consolidated grant ``Particle Physics at the Higgs Centre".

	\bibliographystyle{JHEP}
	\bibliography{references}

\end{document}